\documentclass[10pt,journal,epsfig]{IEEEtran}

\usepackage[dvips]{graphicx}
\usepackage{graphicx}
\usepackage{amssymb}
\usepackage{cite}
\usepackage{subfigure}
\usepackage{amsmath}
\usepackage{footnote}

\IEEEoverridecommandlockouts

\begin{document}

\title{An Efficient Bayesian PAPR Reduction Method for OFDM-Based Massive MIMO Systems}

\author{Hengyao Bao, Jun Fang, Zhi Chen, Hongbin Li,~\IEEEmembership{Senior
Member,~IEEE}, and Shaoqian Li,~\IEEEmembership{Fellow,~IEEE}
\thanks{Hengyao Bao, Jun Fang, Zhi Chen, and Shaoqian Li are with the National Key Laboratory
of Science and Technology on Communications, University of
Electronic Science and Technology of China, Chengdu 611731, China,
Email: JunFang@uestc.edu.cn, chenzhi@uestc.edu.cn,
lsq@uestc.edu.cn}
\thanks{Hongbin Li is
with the Department of Electrical and Computer Engineering,
Stevens Institute of Technology, Hoboken, NJ 07030, USA, E-mail:
Hongbin.Li@stevens.edu}
\thanks{This work was supported in part by the National Science
Foundation of China under Grant 61172114.}}

\maketitle

\begin{abstract}
We consider the problem of peak-to-average power ratio (PAPR)
reduction in orthogonal frequency-division multiplexing (OFDM)
based massive multiple-input multiple-output (MIMO) downlink
systems. Specifically, given a set of symbol vectors to be
transmitted to $K$ users, the problem is to find an OFDM-modulated
signal that has a low PAPR and meanwhile enables multiuser
interference (MUI) cancelation. Unlike previous works that tackled
the problem using convex optimization, we take a Bayesian approach
and develop an efficient PAPR reduction method by exploiting the
redundant degrees-of- freedom of the transmit array. The
sought-after signal is treated as a random vector with a
hierarchical truncated Gaussian mixture prior, which has the
potential to encourage a low PAPR signal with most of its samples
concentrated on the boundaries. A variational
expectation-maximization (EM) strategy is developed to obtain
estimates of the hyperparameters associated with the prior model,
along with the signal. In addition, the generalized approximate
message passing (GAMP) is embedded into the variational EM
framework, which results in a significant reduction in
computational complexity of the proposed algorithm. Simulation
results show our proposed algorithm achieves a substantial
performance improvement over existing methods in terms of both the
PAPR reduction and computational complexity.
\end{abstract}


\section{Introduction}
Massive multiple-input multiple-output (MIMO), also known as
large-scale or very-large MIMO, is a promising technology to meet
the ever growing demands for higher throughput and better
quality-of-service of next-generation wireless communication
systems \cite{RusekPersson13}. Massive MIMO systems are those that
are equipped with a large number of antennas at the base station
(BS) simultaneously serving a much smaller number of
single-antenna users sharing the same time-frequency bandwidth. In
addition to higher throughput, massive MIMO systems also have the
potential to improve the energy efficiency and enable the use of
inexpensive, low-power components. Hence, it is expected that
massive MIMO will bring radical changes to future wireless
communication systems.

In practice, broadband wireless communications may suffer from
frequency-selective fading. Orthogonal frequency-division
multiplexing (OFDM), a scheme of encoding digital data on multiple
carrier frequencies, has been widely used to deal with
frequency-selective fading. However, a major problem associated
with the OFDM is that it is subject to a high peak-to-average
power ratio (PAPR) owing to the independent phases of the
sub-carriers \cite{WunderFischer13}. To avoid out-of-band
radiation and signal distortion, handling this high PAPR requires
a high-resolution digital-to-analog converter (DAC) and a linear
power amplifier (PA) at the transmitter, which is not only
expensive but also power-inefficient \cite{JiangWu08}. The
situation deteriorates when the number of antennas is large,
leaving such systems impractical. Therefore, it is of crucial
importance to reduce the PAPR of massive MIMO-OFDM systems to
facilitate low-cost and power-efficient hardware implementations.

Many techniques have been developed for PAPR reduction in
single-input single-output (SISO) OFDM wireless systems. The most
prominent are clipping \cite{Clipping}, tone reservation (TR)
\cite{TR}, active constellation extension (ACE) \cite{ACE},
selected mapping (SLM) \cite{SLM}, partial transmission sequence
(PTS) \cite{PTS} and others. For a detailed overview, we refer
readers to \cite{HanLee05,JiangWu08}. Although these
PAPR-reduction schemes can be extended to point-to-point MIMO
systems easily \cite{HanLee05,FischerHoch06,TsiJones10}, extension
to the multi-user (MU) MIMO downlink is not straightforward,
mainly because joint receiver-side signal processing is almost
impossible in practice as the users are distributed. Recently, a
new PAPR reduction method \cite{StuderLarsson13} was developed for
massive MIMO-OFDM systems. The proposed scheme utilizes the
redundant degrees-of-freedom (DoFs) resulting from the large
number of antennas at the BS to achieve joint multiuser
interference (MUI) cancelation and PAPR reduction. Specifically,
the problem was formulated as a linear constrained $\ell_\infty$
optimization problem and a fast iterative truncation algorithm
(FITRA) was developed in \cite{StuderLarsson13}. However, the
FITRA algorithm shows to have a fairly low convergence rate. Also,
the algorithm employs a regularization parameter to achieve
balance between the PAPR reduction and the MUI cancelation (i.e.
data fitting error). The choice of the regularization parameter
may be tricky in practice. On the other hand, the regularization
parameter may be seen instead as an additional degree of freedom
that allows to regulate the operation of the algorithm. In
\cite{PrabhuEdfors14}, a peak signal clipping scheme was employed
to reduce the PAPR and some of the antennas at the BS are reserved
to compensate for peak-clipping signals. This method has a lower
computational complexity. But it achieves only a mild PAPR
reduction and those antennas reserved for compensation may incur
large PAPRs.

In this paper, we develop a novel Bayesian approach to address the
joint PAPR reduction and MUI cancelation problem for downlink
multi-user massive MIMO-OFDM systems. Specifically, MUI
cancelation can be formulated as an underdetermined linear inverse
problem which admits numerous solutions. To search for a low PAPR
solution, a hierarchical truncated Gaussian mixture prior model is
proposed and assigned to the unknown signal (i.e. solution). This
hierarchical prior has the potential to encourage a quasi-constant
magnitude solution with as many entries as possible lying on the
truncated boundaries, thus resulting in a low PAPR. A variational
expectation-maximization (EM) algorithm is developed to obtain
estimates of the hyperparameters associated with the prior model,
along with the signal. In addition, the generalized approximate
message passing (GAMP) technique \cite{GAMP} is employed to
facilitate the algorithm development in the expectation step. This
GAMP technique also helps significantly reduce the computational
complexity of the proposed algorithm. Simulation results show that
the proposed method presents a substantial improvement over the
FITRA algorithm in terms of both PAPR reduction and computational
complexity.


During the review process of the current work, it was brought to
our attention that an efficient approximate message passing
(AMP)-based Bayesian method was recently proposed
\cite{ChenWang15} for PAPR reduction for massive MIMO systems,
which can be extended to the case with OFDM modulation. The
rationale behind our work and the above work are similar: both
methods cast the PAPR reduction problem as a Bayesian inference
problem and employ priors to promote solutions with constant
envelopes. The prior distributions employed by these two works,
however, are very different. The prior proposed in
\cite{ChenWang15} assigns each coefficient to a random point on a
circle with a certain radius on the complex plane. Unlike our
work, this prior only encourages entries of the obtained solution
to be close to the boundary but cannot guarantee that they exactly
lie on the boundary points.

The rest of this paper is organized as follows. In Section
\ref{sec:problem-formulation}, we introduce the data model, basic
assumptions, and the PAPR reduction problem. A new hierarchical
Bayesian prior model is proposed in Section
\ref{sec:Bayesian-model}, and an efficient Bayesian algorithm is
developed in Section \ref{sec:algorithm}. Simulation results are
provided in Section \ref{sec:simulation}, followed by concluding
remarks in Section \ref{sec:conclusion}.

\emph{Notations:} Lowercase boldface is used for column vectors
$\boldsymbol{x}$, and uppercase for matrices $\boldsymbol{X}$. The
superscripts $(\cdot)^{T}$ and $(\cdot)^{H}$ represent the
transpose and conjugate transpose, respectively.
$\|\boldsymbol{x}\|_{2}$ is used to denote the $\ell_{2}$ norm of
vector $\boldsymbol{x}$, and $\|\boldsymbol{x}\|_{\infty}$ stands
for the $\ell_{\infty}$ norm, $\ell_{\widetilde{\infty}}$ norm is
define as
$\|\boldsymbol{x}\|_{\ell_{\widetilde{\infty}}}=\max\{\|\Re\{\boldsymbol{x}\}\|_{\infty},
\|\Im\{\boldsymbol{x}\}\|_{\infty}\}$, with
$\Re\{\boldsymbol{x}\}$ and $\Im\{\boldsymbol{x}\}$ denoting the
real and imaginary part of $\boldsymbol{x}$, respectively.
$\boldsymbol{F}_{N}$ denotes the $N\times N$ unitary discrete
Fourier transform (DFT) matrix. The $N\times N$ identity matrix
and the $M\times N$ all-zeros matrix are denoted by
$\boldsymbol{I}_{N}$ and $\boldsymbol{0}_{M\times N}$,
respectively. We denote the pdf of Gaussian random variable $x$
with mean $\mu$ and variance $\sigma^2$ as
$\mathcal{N}(x;\mu,\sigma^2)$, for the special case of
$\mathcal{N}(x;0,1)$, we write the cdf as $\Phi(x)$. The symbol
$\otimes$ denotes the Kronecker product.

\section{System Model and Problem Formulation}
\label{sec:problem-formulation} We first introduce the system
model of OFDM based massive MIMO systems. Then we discuss some
recent research on PAPR reduction for multi-user massive MIMO-OFDM
systems.

\begin{figure*}[!t]
\centering
\includegraphics[width=17.4cm]{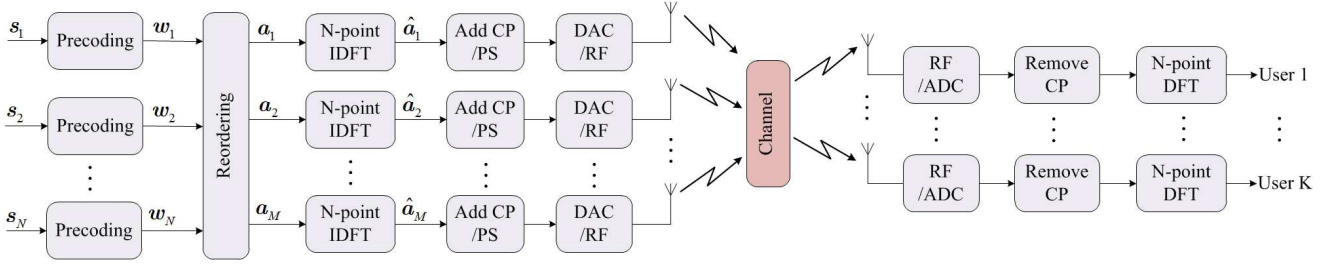}
\caption{System model for the downlink of OFDM based massive MIMO,
with $N$ OFDM tones, $\!M$ transmit antennas and $K$ independent single-antenna users.} \label{systemmodel}
\end{figure*}

\subsection{System Model}
The system model of the OFDM-based massive MIMO downlink scenario
is depicted in Fig. \ref{systemmodel}, where the BS is assumed to
have $M$ transmit antennas and serve $K$ independent
single-antenna users ($K\ll M$), and the total number of OFDM
tones is $N$. In practice, the set of tones available are divided
into two sets $\mathcal{T}$ and $\mathcal{T}^\mathcal{C}$, where
the tones in set $\mathcal{T}$ are used for data transmission and
the tones in its complementary set $\mathcal{T}^\mathcal{C}$ are
used for guard band (unused tones at both ends of the spectrum).
Hence, for each tone $n\in\mathcal{T}$, the corresponding
${K\times1}$ vector $\boldsymbol{s}_{n}$ comprises the symbols for
$K$ users, which are usually chosen from a complex-valued signal
alphabet $\mathcal{B}$. We normalize the data vector to satisfy
$\mathbb{E}\{\|\boldsymbol{s}_{n}\|^{2}_{2}\}=1$. For each tone
$n\in\mathcal{T}^\mathcal{C}$, we set
$\boldsymbol{s}_{n}=\boldsymbol{0}_{K\times1}$ such that no signal
is transmitted in the guard band.

Since cooperative detection among users is often impossible,
precoding must be performed at the BS to remove multi-user
interference (MUI). Usually, the signal vector on the $n$th tone
is linearly precoded as
\begin{align}
\boldsymbol{w}_{n}=\boldsymbol{P}_{n}\boldsymbol{s}_{n},
\label{eqn-1}
\end{align}
where $\boldsymbol{w}_{n}\in\mathbb{C}^{M\times1}$ is the precoded
vector that contains symbols to be transmitted on the $n$th
sub-carrier through the $M$ antennas respectively, and
$\boldsymbol{P}_{n}\in \mathbb{C}^{M\times K}$ represents the
precoding matrix for the $n$th OFDM tone. Zero-forcing (ZF)
precoding and minimum-mean square-error (MMSE) precoding are two
classical precoding schemes. The former aims at removing MUI
completely, while the latter tries to achieve balance between the
MUI cancellation and the noise enhancement. In this paper, we
consider the ZF procoding scheme. Note that since $K\ll M$, the ZF
precoding matrix has an infinite number of forms, among which the
most widely used is
\begin{align}
\boldsymbol{P}_{n}^\mathrm{zf}=\boldsymbol{H}_{n}^{H}(\boldsymbol{H}_{n}\boldsymbol{H}_{n}^{H})^{-1},
\label{ZF}
\end{align}
where $\boldsymbol{H}_{n}\in\mathbb{C}^{K\times M}$ denotes the
MIMO channel matrix associated with the $n$th tone. Here we assume
the channel matrix $\boldsymbol{H}_{n}$, $\forall n$ to be known
at the transmitter, which can be acquired by exploiting the
channel reciprocity of time division duplexing (TDD) systems
(i.e., the downlink channel is the transpose of the uplink
channel).

After precoding, all precoded vectors $\boldsymbol{w}_{n}$ are
reordered to $M$ antennas for OFDM modulation,
\begin{align}
[\boldsymbol{a}_{1}\cdot\cdot\cdot\boldsymbol{a}_{M}]=
[\boldsymbol{w}_{1}\cdot\cdot\cdot\boldsymbol{w}_{N}]^{T},
\label{reorder}
\end{align}
where $\boldsymbol{a}_{m}\in\mathbb{C}^{N\times1}$ represents the
frequency-domain signal to be transmitted from the $m$th antenna.
The time-domain signals are obtained through the inverse discrete
Fourier transform (IDFT), i.e.,
$\boldsymbol{\hat{a}}_{m}=\boldsymbol{F}_{N}^{H}\boldsymbol{a}_{m}$,
$\forall m$. Then, a cyclic prefix (CP) is added to the
time-domain samples of each antenna to eliminate intersymbol
interference (ISI). Finally, these samples are converted to analog
signals and transmitted via the frequency-selective channel.

At the receivers, after removing the CPs of the received signals,
the DFT is performed to obtain the frequency-domain signals. The
receive vector consisting of $K$ users' signals can be described
as
\begin{align}
\boldsymbol{r}_{n}=\boldsymbol{H}_{n}\boldsymbol{w}_{n}+\boldsymbol{e}_{n},
\quad \forall n\label{transmition}
\end{align}
where $\boldsymbol{r}_{n}\in\mathbb{C}^{K\times1}$ denotes the
receive vector associated with the $n$th tone, and
$\boldsymbol{e}_{n}\in\mathbb{C}^{K\times1}$ is the receiver noise
and has i.i.d. circularly symmetric complex Gaussian entries with
zero-mean and variance $N_o$. If the ZF precoding scheme is used,
by combining (\ref{eqn-1}), (\ref{ZF}) and (\ref{transmition}),
the received signal vector equals to
$\boldsymbol{r}_{n}=\boldsymbol{s}_{n}+\boldsymbol{e}_{n}$,
$\forall n$, which means the MUI is perfectly removed.

\subsection{Peak-to-Average Power Ratio (PAPR) Reduction}
OFDM modulation typically exhibits a large dynamic range because
the phases of the sub-carriers are independent of each other,
which may combine in a constructive or destructive manner. To
avoid out-of-band radiation and signal distortion, high-solution
DACs and linear power amplifiers are required at the transmitter
to accommodate the large peaks of OFDM signals, which leads to
expensive and power-inefficient RF chains.

PAPR is defined as the ratio of the peak power of the signal to
its average power. Specifically, the PAPR at the $m$th transmit
antenna is defined as
\begin{align}
\text{PAPR}_{m}\triangleq
\frac{2N\|\boldsymbol{\hat{a}}_{m}\|^2_{\widetilde{\infty}}}{\|\boldsymbol{\hat{a}}_{m}\|_2^2},
\end{align}
where the operator $\|\cdot\|^2_{\widetilde{\infty}}$ is used
because RF-chains often process and modulate the real and
imaginary part of time-domain samples independently. It should
also be noted that, we only consider the PAPR of discrete-time
OFDM signals in this paper, one can obtain its continuous-time
counterpart precisely by implementing an $L$-times oversampling in
OFDM modulation \cite{Tellambura01}.\footnote{Instead of $N$-point
IDFT, $L$-times oversampling can be implemented by $LN$-point IDFT
of the frequency-domain signals with $(L-1)N$ zero-padding.} Since
many conventional MIMO-OFDM systems, such as 3GPP LTE \cite{LTE}
and IEEE 802.11.n \cite{802.11.n}, disallow such an oversampling
operation, here we ignore the difference as in \cite{StuderLarsson13} (i.e., $L=1$).

When the number of transmit antennas is larger than the number of
users, numerous ZF precoding matrices are available. In other
words, for a set of $\boldsymbol{s}_n$, $n=1,...,N$, we have an
infinite number of precoded signals ${\boldsymbol{w}}\triangleq
[\boldsymbol{w}_1^T,...,\boldsymbol{w}_N^T]^T$ that achieve
perfect MUI cancelation. Thus there may exist a candidate
$\boldsymbol{w}$ whose associated time-domain signals
$\{\boldsymbol{\hat{a}}_m\}$ have low PAPRs. In this paper,
instead of designing the procoding matrix, we directly search for
the signal $\boldsymbol{w}$ to achieve a joint PAPR reduction and
MUI cancelation. Specifically, in order to remove the MUI, the
precoded vectors $\boldsymbol{w}_n$ need to satisfy:
\begin{subequations}\label{precoding}
\begin{align}
\boldsymbol{s}_{n}=\boldsymbol{H}_{n}\boldsymbol{w}_{n}, &\ \ n\in \mathcal{T},
\label{precoding_zf}   \\
\boldsymbol{0}_{M\times1}=\boldsymbol{w}_{n}, &\ \ n\in \mathcal{T}^{c}.
\end{align}
\end{subequations}
The whole linear constraints of (\ref{precoding}) can be further
written as
\begin{align}
\overline{\boldsymbol{s}}=\overline{\boldsymbol{H}}{\boldsymbol{w}}
\label{eqn-2}
\end{align}
where $\overline{\boldsymbol{s}}\in\mathbb{C}^{NK\times 1}$
denotes the concatenation of all vectors on the left-hand side of
(\ref{precoding}), $\overline{\boldsymbol{H}}$ is a block diagonal
matrix with its diagonal blocks equal to $\boldsymbol{H}_{n}$ for
$n\in \mathcal{T}$ and $\boldsymbol{I}_{M}$ for
$n\in\mathcal{T}^{c}$. According to (\ref{reorder}), the
reordering operation can be equivalently written as a linear
transformation, i.e.,
\begin{align}
\boldsymbol{a}=\boldsymbol{T}\boldsymbol{w} \label{eqn-3},
\end{align}
where $\boldsymbol{a}=[\boldsymbol{a}_1^T,...,\boldsymbol{a}_M^T]^T$,
$\boldsymbol{T}$ is a permutation matrix that assigns the $M$
entries of each precoded vector to the $M$ antennas respectively.
Recalling
$\boldsymbol{\hat{a}}_{m}=\boldsymbol{F}_{N}^{H}\boldsymbol{a}_{m},\forall
m$, (\ref{eqn-2}) and (\ref{eqn-3}), we have
\begin{align}
\overline{\boldsymbol{s}}=\overline{\boldsymbol{H}}\boldsymbol{T}^T\overline{\boldsymbol{F}}\widehat{\boldsymbol{a}},
\label{linear}
\end{align}
where $\overline{\boldsymbol{F}}\triangleq\boldsymbol{I}_M\otimes
\boldsymbol{F}_N$, and
$\widehat{\boldsymbol{a}}\triangleq[\boldsymbol{\hat{a}}_1^T,...,\boldsymbol{\hat{a}}_M^T]^T$.
Given a symbol vector $\overline{\boldsymbol{s}}$, our goal is to
search for a signal $\widehat{\boldsymbol{a}}$ satisfying the
above equation (\ref{linear}), and meanwhile its sub-vector
$\boldsymbol{\hat{a}}_m$, i.e. the signal to be transmitted at
each antenna, having a low PAPR. This problem can be formulated as
a minimax problem which minimizes the maximum PAPR among all
antennas subject to the linear constraint defined in
(\ref{linear}). Nevertheless, this problem, as indicated in
\cite{StuderLarsson13}, is complex to solve. To circumvent the
difficulty, the minimax problem is replaced by a constrained
optimization which minimizes the $\ell_{\widetilde{\infty}}$ norm
of $\widehat{\boldsymbol{a}}$, a vector formed by aggregating all
time-domain vectors $\{\boldsymbol{\hat{a}}_m\}$
\cite{StuderLarsson13}
\begin{align}
\min{\|\widehat{\boldsymbol{a}}\|_{\widetilde{\infty}}} \quad
\text{subject to
}\overline{\boldsymbol{s}}=\overline{\boldsymbol{H}}\boldsymbol{T}^T
\overline{\boldsymbol{F}}\widehat{\boldsymbol{a}}.
\end{align}
This problem can be further converted into a real-valued form as
follows \cite{StuderLarsson13}
\begin{align}
\min{\|\boldsymbol{x}\|_\infty} \quad
\text{subject to }\boldsymbol{y}=\boldsymbol{A}\boldsymbol{x},
\label{pmp-inf}
\end{align}
where
\begin{align}
\boldsymbol{y}&\triangleq
\begin{bmatrix}
\Re{\{\overline{\boldsymbol{s}}\}} \\
\Im{\{\overline{\boldsymbol{s}}\}}
\end{bmatrix},\
\boldsymbol{x}\triangleq
\begin{bmatrix}
\Re{\{\widehat{\boldsymbol{a}}\}} \\
\Im{\{\widehat{\boldsymbol{a}}\}}
\end{bmatrix},
\nonumber \\
\boldsymbol{A}\triangleq&
\begin{bmatrix}
\Re{\{\overline{\boldsymbol{H}}\boldsymbol{T}^T\overline{\boldsymbol{F}}\}}
&-\Im{\{\overline{\boldsymbol{H}}\boldsymbol{T}^T\overline{\boldsymbol{F}}\}} \\
\Im{\{\overline{\boldsymbol{H}}\boldsymbol{T}^T\overline{\boldsymbol{F}}\}}
&\Re{\{\overline{\boldsymbol{H}}\boldsymbol{T}^T\overline{\boldsymbol{F}}\}}
\end{bmatrix},
\nonumber
\end{align}
and the dimension of $\boldsymbol{A}$ is
$2(|\mathcal{T}|K+|\mathcal{T}^c|M)\times 2NM$. For notational
convenience, let $J\triangleq 2(|\mathcal{T}|K+|\mathcal{T}^c|M)$
and $I\triangleq 2NM$.




Intuitively, via minimizing the largest magnitude of entries of
$\boldsymbol{x}$, the PAPR associated with each transmit antenna
can be reduced. This problem can be solved exactly by
reformulating (\ref{pmp-inf}) as a linear programming problem, but
is computationally prohibitive when the signal dimension is large.
To develop an efficient algorithm, the equality constraint is
relaxed as
$\|\boldsymbol{y}-\boldsymbol{A}\boldsymbol{x}\|_2\leq\delta$ in
\cite{StuderLarsson13}. Hence the optimization (\ref{pmp-inf}) can
eventually be reformulated as
\begin{align}
\min{~\lambda\|\boldsymbol{x}\|_\infty}+\|\boldsymbol{y}-\boldsymbol{A}\boldsymbol{x}\|_2^2,
\label{opt-1}
\end{align}
where $\lambda>0$ is a regularization parameter. An efficient
numerical method, namely, the fast iterative truncation algorithm
(FITRA), was employed \cite{StuderLarsson13} to solve
(\ref{opt-1}). The FITRA algorithm requires to choose a suitable
regularization parameter $\lambda$ to balance between the PAPR
reduction and the data fitting error, which may be tricky in
practice. In the following, we develop a Bayesian method which is
free of this issue, and also turns out to be more efficient and
effective than the FITRA algorithm.

\section{Bayesian Model}
\label{sec:Bayesian-model}
To facilitate our algorithm development, we introduce a noise term
to model the mismatch between $\boldsymbol{y}$ and
$\boldsymbol{A}\boldsymbol{x}$, i.e.
\begin{align}
\boldsymbol{y}=\boldsymbol{A}\boldsymbol{x}+\boldsymbol{\epsilon},
\label{measurement}
\end{align}
where $\boldsymbol{\epsilon}$ denotes the noise vector and its
entries are assumed to be i.i.d. Gaussian random variables with
zero-mean and unknown variance $\beta^{-1}$. Here we treat $\beta$
as an unknown parameter because the Bayesian framework allows an
automatic determination of its model parameters and usually
provides a reasonable balance between the data fitting error and
the desired characteristics of the solution. In case that there is
a pre-specified tolerance value for the MUI, we can also set an
appropriate value for $\beta$ instead of treating it as unknown.

To reduce the PAPR associated with each transmit antenna, we aim
to find a quasi-constant magnitude solution to the above
underdetermined linear system. Note that a constant magnitude
signal achieves a minimum PAPR. Ideally we hope to find a solution
with all of its entries having a constant magnitude. Nevertheless,
it is highly unlikely that there exists such a solution to satisfy
(or approximately satisfy with a tolerable error) the MUI
cancelation equality, i.e. (\ref{measurement}). Therefore we,
alternatively, seek a quasi-constant magnitude solution with as
many entries as possible located on the boundary points of an
interval $[-v,v]$, whereas the rest entries bounded within
$[-v,v]$ but not restricted to lie on the boundary points in order
to meet the MUI cancelation constraint.



To encourage a quasi-constant magnitude solution, we propose a
hierarchical truncated Gaussian mixture prior for the signal
$\boldsymbol{x}$. In the first layer, coefficients of
$\boldsymbol{x}$ are assumed independent of each other and each
entry $x_i$ is assigned a truncated Gaussian mixture distribution:
\begin{align}
p(x_i)=\left\{
\begin{aligned}
\pi\frac{\mathcal{N}(x_i;v,\alpha_{i1}^{-1})}{\eta_{i1}}+
(1-\pi)\frac{\mathcal{N}(x_i;-v,\alpha_{i2}^{-1})}{\eta_{i2}}\\
\text{if } x_i\in[-v,v],\\
0~~~~~~~~~~~~~~~~~~\text{otherwise},~~~~~
\end{aligned}
\right.\label{prior}
\end{align}
where the first component of (\ref{prior}) is characterized by a
truncated Gaussian distribution with its mean and variance given
by $v$ and $\alpha_{i1}^{-1}$, respectively; the second component
is characterized by a truncated Gaussian distribution with its
mean and variance given by $-v$ and $\alpha_{i2}^{-1}$,
respectively; $\pi\in [0,1]$ is a mixing coefficient that denotes
the probability of generating $x_i$ from the first component; the
distribution lies within the interval $[-v,v]$, i.e. from the mean
of the second component to the mean of the first component; and
$\eta_{il}$ is a normalization constant of the $l$th component,
given by
\begin{align}
\eta_{i1}=\frac{1}{2}-\Phi(-2v\sqrt{\alpha_{i1}}),~~  \eta_{i2}=\Phi(2v\sqrt{\alpha_{i2}})-\frac{1}{2}.
\label{normconst}
\end{align}
The second layer specifies Gamma distributions as hyperpriors over
the precision parameters
$\boldsymbol{\alpha}_1\triangleq\{\alpha_{i1}\}_{i=1}^I$ and
$\boldsymbol{\alpha}_2\triangleq\{\alpha_{i2}\}_{i=1}^I$:
\begin{align}
p(\boldsymbol{\alpha}_1,\boldsymbol{\alpha}_2;a,b)=
\prod_{l=1}^2\prod_{i=1}^I \text{Gamma}(\alpha_{li}|a,b),
\end{align}
where
\begin{align}
\text{Gamma}(\alpha|a,b)=\Gamma(a)^{-1}b^{a}\alpha^{a-1}e^{-b\alpha},
\end{align}
in which $\Gamma(a)=\int_{0}^{\infty}t^{a-1}e^{-t}dt$ is the
\textit{gamma} function. To make the hyperpriors non-informative,
small values of $a$ and $b$, e.g. $a=b=10^{-6}$, should be used
\cite{Tipping01}. Note that the choice of the Gamma hyperprior
over the precision is inspired by \cite{Tipping01}. As indicated
in \cite{Tipping01}, the Gamma hyperprior with $a=b=10^{-6}$
corresponds to a broad hyperprior which allows the precision (more
precisely, the posterior mean of the precision) to become
arbitrarily large. For our case, we also place a broad hyperprior
on the precision parameters such that some of these precision
parameters are allowed to become arbitrarily large. As a
consequence, the corresponding entries will be driven towards and
eventually located on the boundary points.



\begin{figure}[!t]
\centering
\setlength{\abovecaptionskip}{0pt}
\setlength{\belowcaptionskip}{0pt}
\includegraphics[width=6.8cm]{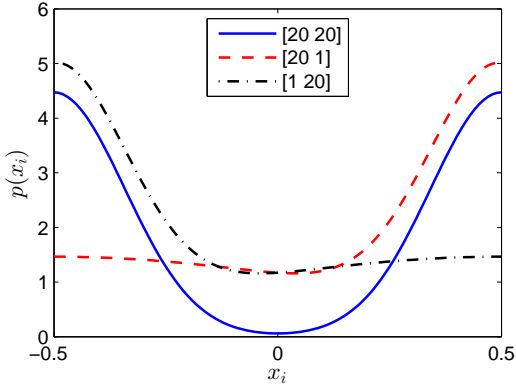}
\caption{Prior distribution function with different $[\alpha_{i1}~~\alpha_{i2}]$,
in which $\pi$ and $v$ are both set to $0.5$.}\label{prior_graph}
\end{figure}

\begin{figure}[!t]
\centering
\setlength{\abovecaptionskip}{0pt}
\setlength{\belowcaptionskip}{0pt}
\includegraphics[width=7cm]{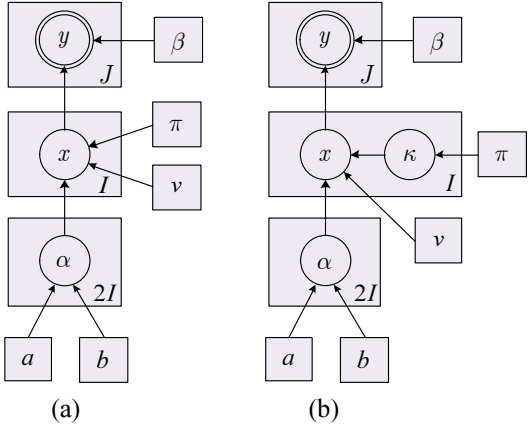}
\caption{Graphical models for low-PAPR signal priors, with circles
denoting hidden variables, double circles denoting observed variables
and squares representing model parameters. (a) Original prior, (b) Modified prior.} \label{graphmodel}
\end{figure}

The prior distributions with different model hyperparameters
$\alpha_{i1}$, $\alpha_{i2}$ are illustrated in Fig.
\ref{prior_graph}, where $\pi$ and $v$ are both set to $0.5$. We
can see that the prior distribution defined in (\ref{prior})
resembles the shape of a bowl. Thus the prior has the potential to
push the entries of the solution toward its boundaries. In
addition, the use of the Gamma hyperprior allows the posterior
mean of the precision to become arbitrarily large. As a result,
the associated entries $x_{i}$ will eventually lie on one of the
two boundary points, leading to a quasi-constant magnitude
solution. The graphical model of the proposed hierarchical is
presented in Fig. \ref{graphmodel}(a).

In general, Bayesian inference requires computing the logarithm
of the prior. In this regard, (\ref{prior}) is a inconvenient form for
inference. To address this issue, we turn the prior into an
exponential form by introducing a binary latent variable
$\kappa_i$ indicating which component is selected for
$x_i$, i.e., $\kappa_i=1$ indicates the first component is
selected while $\kappa_i=0$ corresponds to the second component.
The equivalent prior can be written as
\begin{align}
&p(x_i|\alpha_{i1},\alpha_{i2},\kappa_i;v)\nonumber\\
&\!=\!\left(\frac{\mathcal{N}(x_i;v,\alpha_{i1}^{-1})}{\eta_{i1}}\right)^{\kappa_i}\!
\left(\frac{\mathcal{N}(x_i;-v,\alpha_{i2}^{-1})}{\eta_{i2}}\right)^{1-\kappa_i}\!,x_i\in[-v,v],
\label{prior1}
\end{align}
and the distribution for $\kappa_i$ is
\begin{align}
p(\kappa_i;\pi)=(\pi)^{\kappa_i}(1-\pi)^{1-\kappa_i}.
\label{p(kappa)}
\end{align}
where the mixing coefficient is set to $\pi\!=\!0.5$ to make the
prior non-informative. Also, we define
$\boldsymbol{\kappa}\triangleq \{\kappa_i\}_{i=1}^I$. The updated
graphical model is shown in Fig. \ref{graphmodel}(b). Note that,
according to (\ref{prior1}) and (\ref{p(kappa)}), we can compute
the conditional distribution $p(x_i|\alpha_{i1},\alpha_{i2};v)$
via
$p(x_i|\alpha_{i1},\alpha_{i2};v)\!=\!\sum_{\kappa_i}p(x_i|\alpha_{i1},\alpha_{i2},\kappa_i;v)p(\kappa_i;\pi)$,
which results in the same form of (\ref{prior}).

\section{Bayesian Inference}
\label{sec:algorithm} We now proceed to perform Bayesian inference
for the proposed hierarchical model. A variational
expectation-maximization (EM) strategy is employed for the
Bayesian inference. In our model,
$\boldsymbol{z}\triangleq\{\boldsymbol{x},\boldsymbol{\alpha}_{1},\boldsymbol{\alpha}_{2},\boldsymbol{\kappa}\}$
are treated as hidden variables. The noise variance $\beta$ and
the boundary parameter $v$ are unknown deterministic parameters,
i.e. $\boldsymbol{\theta}\triangleq\{\beta,v\}$. Before
proceeding, we provide a brief review of the variational EM
algorithm.

\subsection{Variational Bayesian Methodology}
Consider a probabilistic model with observed data
$\boldsymbol{y}$, hidden variables $\boldsymbol{z}$ and unknown
deterministic parameters $\boldsymbol{\theta}$. It is
straightforward to show that the marginal probability of the
observed data can be decomposed into two terms
\begin{align}
\ln
p(\boldsymbol{y};\boldsymbol{\theta})=F(q,\boldsymbol{\theta})+\mathrm{KL}(q\|p),\label{log-likelihood}
\end{align}
where
\begin{align}
F(q,\boldsymbol{\theta})=\int
q(\boldsymbol{z})\ln\left(\frac{p(\boldsymbol{y},\boldsymbol{z};
\boldsymbol{\theta})}{q(\boldsymbol{z})}\right)d\boldsymbol{z}
\end{align}
and
\begin{align}
\mathrm{KL}(q\|p)=-\int
q(\boldsymbol{z})\ln\left(\frac{p(\boldsymbol{z}|\boldsymbol{y};
\boldsymbol{\theta})}{q(\boldsymbol{z})}\right)d\boldsymbol{z},
\end{align}
where $q(\boldsymbol{z})$ is any probability density function,
$\mathrm{KL}(q\|p)$ is the Kullback-Leibler divergence between
$p(\boldsymbol{z}|\boldsymbol{y};\boldsymbol{\theta})$ and
$q(\boldsymbol{z})$. Since $\mathrm{KL}(q\|p)\geq0$, it follows
that $F(q,\boldsymbol{\theta})$ is a lower bound of $\ln
p(\boldsymbol{y};\boldsymbol{\theta})$, with the equality holds
only when $\mathrm{KL}(q\|p)=0$, which implies
$p(\boldsymbol{z}|\boldsymbol{y};\boldsymbol{\theta})=q(\boldsymbol{z})$.
The EM algorithm can be viewed as an iterative algorithm which
iteratively maximizes the lower bound $F(q,\boldsymbol{\theta})$
with respect to the distribution $q(\boldsymbol{z})$ and the
parameters $\boldsymbol{\theta}$.

Assume that the current estimate of the parameters is
$\boldsymbol{\theta}^\text{OLD}$. The EM algorithm evaluates
$q^\text{NEW}(\boldsymbol{z})$ by maximizing
$F(q,\boldsymbol{\theta}^\text{OLD})$ with respect to
$q(\boldsymbol{z})$ in the E-step, and then finds new parameter
estimate $\boldsymbol{\theta}^\text{NEW}$ by maximizing
$F(q^\text{NEW},\boldsymbol{\theta})$ with respect to
$\boldsymbol{\theta}$ in the M-step. It is easy to see that when
$q^\text{NEW}(\boldsymbol{z})=p(\boldsymbol{z}|\boldsymbol{y};\boldsymbol{\theta}^\text{OLD})$,
the lower bound $F(q,\boldsymbol{\theta}^\text{OLD})$ is
maximized. Nevertheless, in practice, the posterior distribution
$p(\boldsymbol{z}|\boldsymbol{y};\boldsymbol{\theta}^\text{OLD})$
is usually computationally intractable. To address this
difficulty, we could assume $q(\boldsymbol{z})$ has some specific
parameterized functional form and conduct optimization over the
designated form. A particular form of $q(\boldsymbol{z})$ that has
been widely used with great success is the factorized form over
the component variable or the block component variable $\{z_i\}$
in $\boldsymbol{z}$ \cite{TzikasLikas08}, i.e.
$q(\boldsymbol{z})=\prod_{i}q_i(z_i)$. We therefore can compute
the approximate posterior by finding $q(\boldsymbol{z})$ of the
factorized form that maximizes the lower bound
$F(q,\boldsymbol{\theta}^\text{OLD})$. The maximization can be
conducted in an alternating fashion for each hidden variable,
which leads to \cite{TzikasLikas08}
\begin{align}
q_i(z_i)\propto\exp\big(\langle\ln p(\boldsymbol{y},\boldsymbol{z};\boldsymbol{\theta})\rangle_{k\neq{i}}\big).
\label{general-update}
\end{align}
where $\langle\cdot\rangle_{k\neq i}$ denotes an expectation with
respect to the distributions $q_k(z_k)$ for all $k\neq i$.

Then in the M-step, a new estimate of $\boldsymbol{\theta}$ is
obtained by maximizing the Q-function
\begin{align}
Q(\boldsymbol{\theta},\boldsymbol{\theta}^\text{OLD})=
\left\langle\ln
p(\boldsymbol{y},\boldsymbol{z};\boldsymbol{\theta})\right\rangle
_{q(\boldsymbol{z})}. \label{Q-fun}
\end{align}

\subsection{Likelihood Function Approximation via GAMP}
Let
$\boldsymbol{z}\triangleq\{\boldsymbol{x},\boldsymbol{\alpha}_{1},\boldsymbol{\alpha}_{2},\boldsymbol{\kappa}\}$
denote all hidden variables appearing in our hierarchical model,
and $\boldsymbol{\theta}\triangleq \{\beta,v\}$ denote the unknown
deterministic parameters. As discussed in the previous subsection,
the posterior of $\boldsymbol{z}$ can be approximated by a
factorized form as follows
\begin{align}
p(\boldsymbol{x},\boldsymbol{\alpha}_{1},&\boldsymbol{\alpha}_{2},\boldsymbol{\kappa}|\boldsymbol{y};\beta,v)\nonumber\\
&\approx
q(\boldsymbol{x},\boldsymbol{\alpha}_{1},\boldsymbol{\alpha}_{2},\boldsymbol{\kappa})
=q(\boldsymbol{x})q(\boldsymbol{\alpha}_{1})q(\boldsymbol{\alpha}_{2})q(\boldsymbol{\kappa}).
\end{align}
Following (\ref{general-update}), the approximate posteriors can
be obtained as
\begin{align}
\ln q(\boldsymbol{x})\!=&\langle\ln
p(\boldsymbol{y},\boldsymbol{x},\boldsymbol{\alpha}_{1},\boldsymbol{\alpha}_{2},\boldsymbol{\kappa};\beta,v)
\rangle_{q(\boldsymbol{\alpha}_{1})q(\boldsymbol{\alpha}_{2})q(\boldsymbol{\kappa})}\!+\!\mathrm{const},\nonumber\\
\ln q(\boldsymbol{\alpha}_{1})\!=&\langle\ln p(\boldsymbol{y},\boldsymbol{x},
\boldsymbol{\alpha}_{1},\boldsymbol{\alpha}_{2},\boldsymbol{\kappa};\beta,v)
\rangle_{q(\boldsymbol{x})q(\boldsymbol{\alpha}_{2})q(\boldsymbol{\kappa})}\!+\!\mathrm{const},\nonumber\\
\ln q(\boldsymbol{\alpha}_{2})\!=&\langle\ln
p(\boldsymbol{y},\boldsymbol{x},\boldsymbol{\alpha}_{1},\boldsymbol{\alpha}_{2},
\boldsymbol{\kappa};\beta,v)\rangle_{q(\boldsymbol{x})
q(\boldsymbol{\alpha}_{1})q(\boldsymbol{\kappa})}\!+\!\mathrm{const},\nonumber\\
\ln q(\boldsymbol{\kappa})\!=&\langle\ln
p(\boldsymbol{y},\boldsymbol{x},\boldsymbol{\alpha}_{1},\boldsymbol{\alpha}_{2},
\boldsymbol{\kappa};\beta,v)\rangle_{q(\boldsymbol{x})q(\boldsymbol{\alpha}_{1})q(\boldsymbol{\alpha}_{2})}\!+\!\mathrm{const}.
\end{align}

We first consider the calculation of $q(\boldsymbol{x})$. Keeping
those terms that are dependent on $\boldsymbol{x}$, we have
\begin{align}
&\ln q(\boldsymbol{x})\nonumber\\
&=\left\langle\ln
p(\boldsymbol{y}|\boldsymbol{x};\beta)p(\boldsymbol{x}|
\boldsymbol{\alpha}_{1},\boldsymbol{\alpha}_{2},\boldsymbol{\kappa};v)
\right\rangle_{q(\boldsymbol{\alpha}_{1})q(\boldsymbol{\alpha}_{2})q(\boldsymbol{\kappa})} +\mathrm{const}\nonumber\\
&=\frac{1}{2}\sum_{i=1}^{I}\left\langle-{\alpha_{i1}\kappa_i(x_i-v)^2}-
{\alpha_{i2}(1-\kappa_i)(x_i+v)^2}\right\rangle\nonumber\\
&~~~+\ln p(\boldsymbol{y}|\boldsymbol{x};\beta)+\mathrm{const}
~~~~~~~~~~\text{if }x_i\in[-v,v]~\forall i,\label{lnq(x)}
\end{align}
and $\ln q(\boldsymbol{x})=-\infty$ otherwise, where the
subscripts of $\langle\cdot\rangle_{q(\cdot)}$ are omitted for
simplicity. Since the variables $\{x_i\}$ in the joint likelihood
function $p(\boldsymbol{y}|\boldsymbol{x};\beta)$ are
non-factorizable, obtaining the posterior $q(\boldsymbol{x})$ is
rather difficult. To overcome this difficulty, we employ the
generalized approximate message passing (GAMP) technique
\cite{GAMP} to obtain an amiable approximation of the joint
likelihood function $p(\boldsymbol{y}|\boldsymbol{x};\beta)$.

\begin{figure}[!t]
\centering
\includegraphics[width=8.5cm]{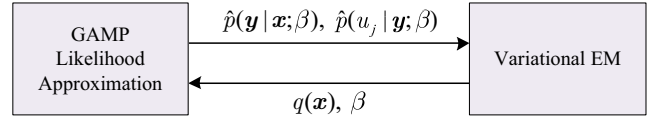}
\caption{Proposed variational EM-GAMP framework, where the hatted distribution $\hat{p}(\cdot)$
represents an approximation of $p(\cdot)$.} \label{GAMP-VEM}
\end{figure}

GAMP is a simplification of loopy BP, and can be used to compute
approximate marginal posteriors and likelihoods. Here we
approximate the joint likelihood function
$p(\boldsymbol{y}|\boldsymbol{x};\beta)$ as a product of
approximate marginal likelihoods computed via the GAMP, i.e.
\begin{align}
p(\boldsymbol{y}|\boldsymbol{x};\beta)\approx\hat{p}(\boldsymbol{y}|\boldsymbol{x};\beta)
\propto\prod_{i=1}^I\mathcal{N}(x_i|\hat{r}_i,\tau_i^r),
\label{likelihood-approximation}
\end{align}
where $\mathcal{N}(x_i|\hat{r}_i,\tau_i^r)$ is the approximate marginal likelihood obtained by the
GAMP algorithm. To calculate $\hat{r}_i$ and $\tau_i^r$, an
estimate of the posterior $q(\boldsymbol{x})$ and $\beta$ is
required as inputs to the GAMP algorithm (see the details of the
GAMP algorithm provided below). Hence the GAMP algorithm can be
embedded in the variational EM framework: given an estimate of
$q(\boldsymbol{x})$ and $\beta$, use the GAMP to obtain an
approximation of the likelihood function
${p}(\boldsymbol{y}|\boldsymbol{x};\beta)$; with the
approximation $\hat{p}(\boldsymbol{y}|\boldsymbol{x};\beta)$, the
variational EM proceeds to yield a new estimate of
$q(\boldsymbol{x})$ and $\beta$, along with estimates of other
deterministic parameters (e.g. $v$) and posterior distributions
for the other hidden variables (e.g.
$\boldsymbol{\alpha}_{1},\boldsymbol{\alpha}_{2},\boldsymbol{\kappa}$).
This iterative procedure is illustrated in Fig. \ref{GAMP-VEM}.

Note that besides the approximation
$\hat{p}(\boldsymbol{y}|x_i;\beta)$, GAMP also produces
approximations for the marginal posteriors of the noiseless output
$\boldsymbol{u}=[u_1,...,u_J]^T\triangleq\boldsymbol{A}\boldsymbol{x}$,
which are given by
\begin{align}
p(u_j|\boldsymbol{y},\beta)&\approx
\hat{p}(u_j|\boldsymbol{y},\beta)\nonumber\\
&\propto
p(y_j|u_j;\beta)\mathcal{N}(u_j|\hat{p}_j,\tau_j^p),
\end{align}
where $\hat{p}_j$ and $\tau_j^p$ are quantities obtained from the
GAMP algorithm. Since the noise is assumed to be white Gaussian noise,
we have
$\hat{p}(u_j|\boldsymbol{y},\beta)=\mathcal{N}(u_j|\widehat{u}_j,\tau_j^z)$,
where
\begin{align}
\tau_j^u=\frac{\tau_j^p}{\tau_j^p\beta+1} \qquad
\widehat{u}_j=\tau_j^u\left({y_j}{\beta}+\frac{\widehat{p}_j}{\tau_j^p}\right).
\end{align}
As will be shown later, this approximation can be
used to learn the inverse of the noise variance, $\beta$, in the
M-step.

\begin{table}
\normalsize
\small
\doublerulesep=0.4pt \vspace{0cm} \noindent
\begin{tabular}{p{8.45cm}}
\hline\hline \textbf{Algorithm 1~~Likelihood Approximation via GAMP }\\
\hline \textbf{Input:} means and variances of posteriors $q(x_i)$:
$\widehat{x}_i=\langle{x}_i\rangle_{q(x_i)}$,
$\tau_i^{x}=\langle{x}_i\rangle^\mathbb{V}_{q(x_i)}$,
$i=1,...,I$, where $\langle\cdot\rangle_{q(\cdot)}^{\mathbb{V}}$ denotes the variance
with respect to ${q(\cdot)}$,
and inverse noise variance $\beta$. Initialize $\widehat{s}_j$ as $0$, $j=1,...,J$.
\\
\textbf{Output:} approximate likelihoods
$\mathcal{N}(x_i|\widehat{r}_i,\tau_i^r)$, $i=1,...,I$,  and
posteriors of $u_j$: $\mathcal{N}(u_j|\widehat{u}_j,\tau_j^u)$,
$j=1,...,J$.
\\

Step $1$. For each $j$: \\
$\begin{aligned}
~~~~~~~~~~~~~~~~~~~\tau_j^p=&\sum_{i}A_{ji}^2\tau_i^x \\
                   \widehat{p}_j=&\sum_{i}A_{ji}\widehat{x}_i-\tau_j^p\widehat{s}_j\\
\end{aligned}$\\

Step $2$. For each $j$: \\
$\begin{aligned}
~~~~~~~~~~~~~~~~~~~\widehat{u}_j=&\langle{u_j}\rangle_{p\left(u_j|{y}_j,\widehat{p}_j,\tau_j^p\right)}\\
                   \tau_j^u=&\langle{u_j}\rangle^\mathbb{V}_{p\left(u_j|{y}_j,\widehat{p}_j,\tau_j^p\right)}\\
                   \widehat{s}_j=&\frac{\widehat{u}_j-\widehat{p}_j}{\tau_j^p}\\
                    \tau_j^s=&\frac{1}{\tau_j^p}\left(1-\frac{\tau_j^u}{\tau_j^p}\right)
\end{aligned}$ \\

Step $3$. For each $i$: \\
$\begin{aligned}
~~~~~~~~~~~~~~~~~~~\tau_i^r=&\left(\sum_{j}A_{ji}^2\tau^s_j\right)^{-1} \\
                   \widehat{r}_i=&\widehat{x}_i+\tau_i^r\sum_{j}A_{ji}\widehat{s}_j\\
\end{aligned}$ \\

\hline\hline
\end{tabular}
\end{table}

\emph{Remarks:} Generalized approximate message passing (GAMP) is
a very-low-complexity Bayesian iterative technique recently
developed in \cite{GAMP} for obtaining approximate marginal
posteriors and likelihoods. It therefore can be naturally embedded
within the EM framework to provide an approximate posterior
distribution of $\boldsymbol{x}$ and reduce the computational
complexity, as shown in \cite{EMGMAMP,Non-EMGAMP}. Specifically,
the EM-GAMP framework of \cite{EMGMAMP,Non-EMGAMP} proceeds in a
double-loop manner: the outer loop (EM) computes the Q-function
using the approximate posterior distribution of $\boldsymbol{x}$,
and maximizes the Q-function to update the model parameters (e.g.
$\boldsymbol{\alpha}_{1},\boldsymbol{\alpha}_{2},\boldsymbol{\kappa}$);
the inner loop (GAMP) utilizes the newly estimated parameters to
obtain a new approximation of the posterior distribution of
$\boldsymbol{x}$. However, this procedure is not suitable for our
variational EM framework, because from the GAMP's point of view,
the hyperparameters
$\{\boldsymbol{\alpha}_{1},\boldsymbol{\alpha}_{2},\boldsymbol{\kappa}\}$
need to be known and fixed in order to compute an approximate
posterior distribution of $\boldsymbol{x}$, while the variational
EM treats the model parameters (e.g.
$\boldsymbol{\alpha}_{1},\boldsymbol{\alpha}_{2},\boldsymbol{\kappa}$)
as latent variables. Therefore, instead of computing the
approximate posterior distribution of $\boldsymbol{x}$, in our
variational EM framework, the GAMP is simply used to obtain an
amiable approximation of the likelihood function
$p(\boldsymbol{y}|\boldsymbol{x};\beta)$, and this approximation
involves no latent variables
$\{\boldsymbol{\alpha}_{1},\boldsymbol{\alpha}_{2},\boldsymbol{\kappa}\}$.
Besides, unlike the EM-GAMP framework where the inner loop (GAMP)
is implemented in an iterative way, in our proposed variational
EM-GAMP framework, as detailed in Algorithm 1, the GAMP only needs
to go through one iteration to obtain an approximation of the
likelihood function. In fact, the GAMP algorithm described here is
a simplified version of the original GAMP algorithm by retaining
only its first three steps and skipping its iterative procedure.
Note that the original GAMP algorithm involves a four-step
iterative process, in which the fourth step computes the posterior
of $\boldsymbol{x}$ by using the approximate likelihood function
obtained from the first three steps.

Note that we can also treat
$\{\boldsymbol{\alpha}_{1},\boldsymbol{\alpha}_{2},\boldsymbol{\kappa}\}$
as deterministic parameters and resort to the EM-GAMP framework
for Bayesian inference. Nevertheless, in this case, we need to
estimate a set of binary parameters $\{\kappa_i\}$ in the M-step.
This is essentially a combinatorial search problem and the binary
estimation may cause the algorithm to get stuck in undesirable
local minima.

GAMP is known to work well for $\boldsymbol{A}$ with i.i.d
zero-mean sub-Gaussian entries, but may fail for a rank-deficient
$\boldsymbol{A}$. One may refer to the method \cite{VilaPhilip15}
to improve the stability of the GAMP against the ill-condition of
the matrix $\boldsymbol{A}$. Nevertheless, GAMP is expected to
perform well in wireless communication scenarios since indoor and
urban outdoor environments are typically rich in scattering and
entries of MIMO channel matrices are usually assumed to be i.i.d
Gaussian \cite{ChenWang15,GuoHuang13}.


\subsection{E-Step: Update of Hidden Variables}\label{IV_D}
\textbf{Update of $q(\boldsymbol{x})$:} As discussed above,
$p(\boldsymbol{y}|\boldsymbol{x};\beta)$ is approximated as a
factorized form of $I$ independent scalar likelihoods, which enables
the computation of $q(\boldsymbol{x})$ (\ref{lnq(x)}).
Specifically, using
(\ref{likelihood-approximation}), (\ref{lnq(x)}) can be simplified
as
\begin{align}
&\ln q(\boldsymbol{x})\nonumber\\
&=\frac{1}{2}\sum_{i=1}^{I}\left\langle-{\alpha_{i1}\kappa_i(x_i-v)^2}-
{\alpha_{i2}(1-\kappa_i)(x_i+v)^2}\right\rangle\nonumber\\
&~~~-\frac{1}{2}\sum_{i=1}^{I}{(x_i-\widehat{r}_i)^2}\big/{\tau_i^r}+\mathrm{const}\nonumber\\
&=-\sum_{i=1}^{I}\bigg(\frac12\left(\langle\kappa_i\rangle\langle\alpha_{i1}\rangle-
\langle\kappa_i\rangle\langle\alpha_{i2}\rangle+\langle\alpha_{i2}\rangle+1/\tau_i^r\right)x_i^2\nonumber\\
&~~~+\!\big(\left(\langle\kappa_i\rangle\langle\alpha_{i1}\rangle+
\langle\kappa_i\rangle\langle\alpha_{i2}\rangle-\langle\alpha_{i2}
\rangle\right)v+\widehat{r}_i/\tau_i^r\big)x_i\bigg)\!+\!\mathrm{const}\nonumber\\
&~~~~~~~~~~~~~~~~~~~~~~~~~~~~~~~~~~~~~~~~~\text{if }x_i\in[-v,v]~\forall i,
\end{align}
and $\ln q(\boldsymbol{x})=-\infty$ otherwise. It can be seen that
$\ln q(\boldsymbol{x})$ has a factorized form, which implies that
hidden variables $\{x_i\}$ have independent posterior
distributions. Also, it can be readily verified that the posterior
$q(x_i)$ follows a truncated Gaussian distribution
\begin{align}
q(x_i)=\left\{
\begin{aligned}
\frac{\mathcal{N}(x_i|\mu_i,\sigma_i^2)}{\phi_i} \quad
\text{if }x_i\in[-v,v],\\
0~~~~~~~~~~\text{otherwise},~~~~~
\end{aligned}\right.
\label{q(x)}
\end{align}
where the variance $\sigma_i^2$, mean $\mu_i$ and the
normalization constant $\phi_i$ are given respectively as
\begin{align}
\sigma_i^2&=\left(\langle\kappa_i\rangle\langle\alpha_{i1}\rangle-\langle\kappa_i
\rangle\langle\alpha_{i2}\rangle+\langle\alpha_{i2}\rangle+1/\tau_i^r\right)^{-1},\label{variance_x}\\
\mu_i&=\big{(}\left(\langle\kappa_i\rangle\langle\alpha_{i1}\rangle\!+\langle
\kappa_i\rangle\langle\alpha_{i2}\rangle-\langle\alpha_{i2}\rangle\right)v+
\widehat{r}_i/\tau_i^r\big{)}\sigma_i^2,\label{mean_x}\\
\phi_i&=\Phi\left((v-\mu_i)/\sigma_i\right)-\Phi\left((-v-\mu_i)/\sigma_i\right).
\end{align}

\textbf{Update of $q(\boldsymbol{\alpha}_1)$:} Keeping only the
terms that depend on $\boldsymbol{\alpha}_1$, the variational
optimization of $q(\boldsymbol{\alpha}_{1})$ yields
\begin{align}
&\ln q(\boldsymbol{\alpha}_{1})\nonumber\\
&=\langle \ln p(\boldsymbol{x}|\boldsymbol{\alpha}_{1},\boldsymbol{\alpha}_{2},
\boldsymbol{\kappa};v)p(\boldsymbol{\alpha}_{1})\rangle_{q(\boldsymbol{x})
q(\boldsymbol{\alpha}_{2})q(\boldsymbol{\kappa})}+\mathrm{const}\nonumber\\
&=\sum_{i=1}^{I}\langle\ln
p(x_i|\alpha_{i1},\alpha_{i2},\kappa_i;v)p(\alpha_{i1})\rangle_{q(\boldsymbol{x})
q(\boldsymbol{\alpha}_{2})q(\boldsymbol{\kappa})}+
\mathrm{const}\nonumber\\
&=-\sum_{i=1}^{I}\langle\kappa_i\rangle\ln\eta_{i1}+\sum_{i=1}^{I}
\Bigg(\left(a+\frac12\langle{\kappa_i}\rangle-1\right)\ln\alpha_{i1}\nonumber\\
&~~~-\left(b+\frac12\langle{\kappa_i}\rangle\left\langle(x_i-v)^2\right\rangle\right)\alpha_{i1}\Bigg)+\mathrm{const}.
\end{align}
We see that $\ln q(\boldsymbol{\alpha}_{1})$ also has a factorized
form $\ln q(\boldsymbol{\alpha}_{1})=\sum_i\ln q(\alpha_{i1})$.
Note that $\eta_{i1}$ (defined in (\ref{normconst})) is a function
of $\alpha_{i1}$, which makes the inference of $q(\alpha_{i1})$
difficult. To address this difficulty, we use the latest computed
value to replace $\eta_{i1}$ i.e. let
$\ln\eta_{i1}\approx\ln\eta_{i1}^{(t)}$. Note that similar
approximations were also adopted in \cite{Non-EMGAMP} to
facilitate the inference. With this approximation, we obtain
\begin{align}
&\ln q(\alpha_{i1})\nonumber\\
&=\left(a+\frac12\langle{\kappa_i}\rangle-1\right)\ln\alpha_{i1}-
\left(b+\frac12\langle{\kappa_i}\rangle\left\langle(x_i-v)^2\right\rangle\right)\alpha_{i1}\nonumber\\
&~~~+\mathrm{const}. \label{q(alpha)}
\end{align}
Therefore $q(\alpha_{i1})$ follows a Gamma distribution
\begin{align}
q(\alpha_{i1})=\text{Gamma}(\alpha_{i1}|\widetilde{a}_{i1},\widetilde{b}_{i1})
\end{align}
with
\begin{align}
\widetilde{a}_{i1}&=a+\frac12\langle\kappa_i\rangle\\
\widetilde{b}_{i1}&=b+\frac12\langle\kappa_i\rangle\left\langle(x_i-v)^2\right\rangle.
\end{align}

\textbf{Update of $q(\boldsymbol{\alpha}_2)$:}  Following a
procedure similar to the derivation of $q(\boldsymbol{\alpha}_1)$,
we have
\begin{align}
q(\alpha_{i2})=\text{Gamma}(\alpha_{i2}|\widetilde{a}_{i2},\widetilde{b}_{i2})
\end{align}
with
\begin{align}
\widetilde{a}_{i2}&=a+\frac12(1-\langle\kappa_i\rangle)\\
\widetilde{b}_{i2}&=b+\frac12(1-\langle\kappa_i\rangle)\left\langle(x_i+v)^2\right\rangle.
\end{align}

\textbf{Update of $q(\boldsymbol{\kappa})$:} The approximate
posterior distribution $q_{\kappa}(\boldsymbol{\kappa})$ can be
computed as
\begin{align}
&\ln q(\boldsymbol{\kappa})\nonumber\\
&=\langle\ln p(\boldsymbol{x}|\boldsymbol{\alpha}_{1},\boldsymbol{\alpha}_{2},\boldsymbol{\kappa};v)
p(\boldsymbol{\kappa})\rangle_{q(\boldsymbol{x})q(\boldsymbol{\alpha}_{1})q(\boldsymbol{\alpha}_{2})}+\mathrm{const}\nonumber\\
&=\sum_{i=1}^{I}\langle\ln
p(x_i|\alpha_{i1},\alpha_{i2},\kappa_i;v)p(\kappa_i)\rangle_{q(x_{i})q(\alpha_{i1})q(\alpha_{i2})}+
\mathrm{const}\nonumber\\
&=\sum_{i=1}^{I}\Bigg(\frac12\left(\langle\ln\alpha_{i1}\rangle-\langle\ln\alpha_{i2}\rangle-
\left\langle(x_i-v)^2\right\rangle+\left\langle(x_i+v)^2\right\rangle\right)\nonumber\\
&~~~+\langle\ln\eta_{i2}\rangle-\langle\ln\eta_{i1}\rangle+\ln\frac{\pi}{1-\pi}\Bigg)\kappa_i+\mathrm{const}.
\label{lnq(k)}
\end{align}
We see that $\ln q(\boldsymbol{\kappa})= \sum_i\ln q(\kappa_i)$
and, moreover, the posterior $q(\kappa_i)$ obeys a Bernoulli distribution,
i.e. $\kappa_i$ takes values zero or one, and the corresponding
probability can be computed from (\ref{lnq(k)}). To simplify
computation, we can use the approximation
$\langle\ln\eta_{il}\rangle\approx\ln\eta_{il}^{(t)}, l=1,2$.

In summary, the variational Bayesian inference involves updates of
the approximate posterior distributions for hidden variables
$\boldsymbol{x}$, $\boldsymbol{\alpha}_1$, $\boldsymbol{\alpha}_2$
and $\boldsymbol{\kappa}$ in an alternating fashion. Some of the
expectations and moments used during the update are summarized as
\begin{align}
&\langle{x_i}\rangle=\mu_i-\frac{\sigma_i^2}{\phi_i}{\left(\mathcal{N}(v|\mu_i,\sigma_i^2)-
\mathcal{N}(-v|\mu_i,\sigma_i^2)\right)},\label{E[x]}\\
&\langle{x_i^2}\rangle=u_i\langle{x_i}\rangle\!+\!\sigma_i^2\!-\!\frac{\sigma_i^2}{\phi_i}{
\left(\mathcal{N}(v|\mu_i,\sigma_i^2)\!+\!\mathcal{N}(\!-\!v|\mu_i,\sigma_i^2)\right)},\label{E[x^2]}\\
&\langle\alpha_{il}\rangle=\widetilde{a}_{il}/\widetilde{b}_{il},~~l=1,2,\label{E[alpha]}\\
&\langle\ln\alpha_{il}\rangle=\psi(\widetilde{a}_{il})-\ln\widetilde{b}_{il},~~l=1,2,\\
&\langle\kappa_i\rangle=q(\kappa_i=1), \label{E-step-last}
\end{align}
where
\begin{align}
\psi(a)\triangleq\frac{\partial\ln\Gamma(a)}{\partial a}
\end{align}
is known as the \textit{digamma} function
\cite{abramowitz1964handbook}.

\emph{Discussions:} We can gain some insight into our proposed
algorithm by examining the update rules for precision parameters
$\{\alpha_{i1},\alpha_{i2}\}$. Since $a$ and $b$ are set very
small, the update rules (\ref{E[alpha]}) for
$\{\alpha_{i1},\alpha_{i2}\}$ are approximately given by
\begin{align}
\langle\alpha_{i1}\rangle=&\frac{1}{\langle(x_i-v)^2\rangle}\\
\langle\alpha_{i2}\rangle=&\frac{1}{\langle(x_i+v)^2\rangle}
\end{align}
We see that the posterior mean of the precision, say
$\langle\alpha_{i1}\rangle$, is inversely proportional to the
distance between the entry and the boundary point $v$. When $x_i$
is close to the boundary point $v$, the posterior mean of the
precision $\alpha_{i1}$ will become large. As a consequence, the
prior distribution becomes sharp around the boundary point $v$.
Hence the prior has the potential to push the entry $x_i$ closer
to the boundary point $v$, which in turn results in a larger
$\langle\alpha_{i1}\rangle$ according to (\ref{E[alpha]}). This
feedback mechanism keeps pushing most of the entries towards the
boundary until they are eventually located on one of the boundary
points. Our simulation results further corroborate our above
discussions: the proposed algorithm yields a solution with a
substantial percentage of entries lying exactly on the boundary
points.


\subsection{M-Step: Update of Deterministic Parameters}
As indicated earlier, in the variational EM framework, the
deterministic parameters $\boldsymbol{\theta}=\{\beta,v\}$ are
estimated by maximizing the Q-function, i.e.
\begin{align}
\boldsymbol{\theta}^\text{NEW}=\max_{\boldsymbol{\theta}}\quad
Q(\boldsymbol{\theta},\boldsymbol{\theta}^\text{OLD})=
\left\langle\ln
p(\boldsymbol{y},\boldsymbol{z};\boldsymbol{\theta})\right\rangle
_{q(\boldsymbol{z})}.
\end{align}

\textbf{Update of $\beta$:} We fist discuss the update of the
parameter $\beta$, the inverse of the noise variance. Since the
GAMP algorithm provides an approximate posterior distribution for
the noiseless output $\boldsymbol{u}\triangleq\boldsymbol{Ax}$, we
can simply treat $\boldsymbol{u}$ as hidden variables when
computing the Q-function, i.e.
\begin{align}
Q(\beta,\beta^{(t)})&=\sum_{j=1}^J\langle\ln{p(y_j|u_j;\beta)}\rangle_{\hat{p}(u_j|\boldsymbol{y},\beta)}+
\mathrm{const}\nonumber\\
&=\frac{J}{2}\ln\beta-\frac12\beta\sum_{j=1}^J\left\langle(y_j-u_j)^2\right\rangle+\mathrm{const}.
\end{align}
The new estimate of $\beta$ is obtained by maximizing the
Q-function, which can be solved by setting the derivative of
$Q(\beta,\beta^{(t)})$ with respect to $\beta$ to zero. The
derivative is given as
\begin{align}
\frac{\partial
Q(\beta,\beta^{(t)})}{\partial\beta}&=\frac{J}{2\beta}-\frac12\sum_{j=1}^J\left\langle(y_j-u_j)^2\right\rangle.
\end{align}
Setting it to zero, we obtain
\begin{align}
\beta^{(t+1)}=\frac{J}{\sum_{j=1}^J\left\langle(y_j-u_j)^2\right\rangle}.
\label{update_beta}
\end{align}

\textbf{Update of $v$:} We now discuss how to update the boundary
parameter $v$. The boundary parameter $v$ can be updated by
maximizing the Q-function with respect to $v$. Nevertheless, the
optimization is complex since the Q-function involves computing
the expectation of the normalization terms $\eta_{il}$,
$i=1,...,I$, $l=1,2$, with respect to the posterior distributions
$p(\alpha_{il})$. Here we propose a heuristic approach to update
$v$. The basic idea is to find an appropriate value of $v$ such
that the mismatch
$\|\boldsymbol{y}-\boldsymbol{A}\boldsymbol{\hat{x}}\|_2^2$ is
minimized, where $\boldsymbol{\hat{x}}$ denotes the estimated
signal which is chosen as the mean of the posterior distribution
$q(\boldsymbol{x})$. Note that when the boundary parameter $v$ is
small, the mismatch could be large since there may not exist a
solution to satisfy the constraint
$\boldsymbol{y}=\boldsymbol{A}\boldsymbol{x}$ given that
$\|\boldsymbol{x}\|_{\infty}\leq v$. Therefore we can firstly set
a small value of $v$, then gradually increase $v$ by a step-size
such that the mismatch keeps decreasing and eventually becomes
negligible. Define
$\delta(\boldsymbol{\hat{x}})\triangleq\|\boldsymbol{y}-\boldsymbol{A}\boldsymbol{\hat{x}}\|_2^2$.
Specifically, the step-size $\Delta{v}$ can be obtained by solving
the following optimization problem:
\begin{align}
\Delta{v}=\min_{\Delta{v}}\delta(\boldsymbol{\hat{x}}^{(t)}+\boldsymbol{\gamma}\Delta{v}),
\label{Delta-v}
\end{align}
where $\boldsymbol{\hat{x}}^{(t)}$ denotes the estimate (i.e.
posterior mean of $q(\boldsymbol{x})$) of the signal at iteration
$t$, and $\boldsymbol{\gamma}\triangleq[\gamma_1,...,\gamma_I]^T$
is defined as
\begin{align}
\gamma_i=\left\{
\begin{aligned}
1&,~\text{if}~\hat{x}_i^{(t)}\geq0 \\
-1&,~\text{if}~\hat{x}_i^{(t)}<0
\end{aligned}
\right.\label{update_v1}.
\end{align}
The rationale behind the optimization (\ref{Delta-v}) can be
explained as follows. Since our proposed algorithm yields a
solution with most of its entries located on the boundary points,
if we increase the boundary $v$ by a sufficiently small step-size
$\Delta{v}$, we can expect that the signal $\boldsymbol{x}$ will
expand accordingly. We wish to find a step-size $\Delta{v}$ such
that the expanded signal will result in a reduced mismatch. The
problem (\ref{Delta-v}) is a scalar least-square problem, and its
solution is given by
\begin{align}
\Delta{v}=\frac{(\boldsymbol{y}-\boldsymbol{A}\boldsymbol{\hat{x}}^{(t)})^T
\boldsymbol{A}\boldsymbol{\gamma}}{\|\boldsymbol{A}\boldsymbol{\gamma}\|^2_2}.
\label{update_v2}
\end{align}
Then $v$ can be updated as
\begin{align}
v^{(t+1)}=v^{(t)}+\Delta{v}.\label{update_v3}
\end{align}

\subsection{Summary}
In summary, our algorithm is developed by resorting to the
variational EM strategy. The GAMP technique is embedded in the
variational EM framework to obtain an approximation of the joint
likelihood function $p(\boldsymbol{y}|\boldsymbol{x},\beta)$ which
has a factorized form in terms of the variables $\{x_i\}$.
Specifically, the algorithm involves an iterative process as
follows: given an estimate of $q(\boldsymbol{x})$ and $\beta$, we
use the GAMP to obtain an approximation of the likelihood function
${p}(\boldsymbol{y}|\boldsymbol{x};\beta)$; with the
approximation $\hat{p}(\boldsymbol{y}|\boldsymbol{x};\beta)$, the
variational EM proceeds to yield a new estimate of
$q(\boldsymbol{x})$ and $\beta$, along with the approximate
posteriors of the other hidden variables and an estimate of the
boundary parameter $v$. For clarity, we summarize our proposed
algorithm as follows.

\begin{table}[h]
\small
\doublerulesep=0.4pt \noindent
\begin{tabular}{p{8.45cm}}
\hline\hline \textbf{Algorithm 2~~EM-TGM-GAMP}\\
\hline \textbf{Initialization}: $\beta^{(0)}=10^{3}$,
$v^{(0)}=\|\boldsymbol{y}\|_{\infty}/\|\boldsymbol{A}\|_{\infty}$,
initialize the means of $q(\boldsymbol{x})$,
$q(\boldsymbol{\alpha}_1)$, $q(\boldsymbol{\alpha}_2)$,
$q(\boldsymbol{\kappa})$ as $\boldsymbol{0}$, $\boldsymbol{1}$,
$\boldsymbol{1}$, $\frac{1}{2}\boldsymbol{1}$ respectively, set
the variance of $q(\boldsymbol{x})$ as $\boldsymbol{1}$, and set
iteration number $t=0$.
\\
\textbf{Repeat} the following steps until $t\geq t_{\text{MAX}}$\\

\quad $1$. Based on the mean and variance of $q(\boldsymbol{x})$ and $\beta^{(t)}$,\\
~~~~~~calculate the approximate distributions
$\hat{p}(\boldsymbol{y}|\boldsymbol{x};\beta^{(t)})$ \\
~~~~~~and $\hat{p}(u_j|\boldsymbol{y},\beta^{(t)})$, $j=1,...,J$,
via Algorithm $1$.\\

\quad $2$. Using the approximate likelihood $\hat{p}(\boldsymbol{y}|\boldsymbol{x};\beta^{(t)})$, update\\
~~~~~~the posteriors of hidden variables: $q(\boldsymbol{x})$, $q(\boldsymbol{\alpha}_1)$, $q(\boldsymbol{\alpha}_2)$ \\
~~~~~~and $q(\boldsymbol{\kappa})$ via (\ref{q(x)})-(\ref{E-step-last}).\\

\quad $3$. Compute the new estimate $\beta^{(t+1)}$ using (\ref{update_beta}), and \\
~~~~~~obtain the $v^{(t+1)}$ via (\ref{update_v1})-(\ref{update_v3}).

\quad$4$. Increase $t=t+1$ and return to step $1$.\\
\hline\hline
\end{tabular}
\end{table}

Note that the dominating operations of the proposed algorithm in
each iteration only involve simple matrix-vector multiplications,
which scales as $\mathcal{O}(JI)$ ($J<I$). Thus the proposed
algorithm has a computational complexity comparable to the FITRA
algorithm \cite{StuderLarsson13} which also has a computational
complexity of $\mathcal{O}(JI)$ per iteration. Besides, as will be
shown in our experiments, the proposed algorithm has a much faster
convergence rate than the FITRA algorithm, which is more favorable
for real-time implementation needed for practical systems.

\section{Simulation Results} \label{sec:simulation}
We now carry out experiments to illustrate the effectiveness of
the proposed truncated Gaussian mixture (TGM) model-based
variational EM-GAMP algorithm\footnote{Codes are available at
http://www.junfang-uestc.net/codes/EM-TGM-GAMP.rar} (referred to
as the EM-TGM-GAMP). We compare our approach with the FITRA
algorithm \cite{StuderLarsson13}, the zero-forcing (ZF) precoding
scheme, and the amplitude clipping scheme \cite{Clipping} in which
the ZF is first employed and then the peaks of the resulting
signal are clipped with a specified threshold.

In our simulations, we consider a MIMO system which has $M=100$
antennas at the BS and serves $K=10$ single-antenna users. A
$16$-QAM constellation is considered, and the number of OFDM tones
is set to $N=128$, in which only $|\mathcal{T}|=114$ tones are
used for data transmission \cite{802.11.n}. We assume that the
channel is frequency-selective and modeled as a tap-delay line
with $D=8$ taps. The time-domain channel response matrices
$\widehat{\boldsymbol{H}}_{d}$, $d=1,...,D$, have i.i.d.
circularly symmetric Gaussian distributed entries with zero mean
and unit variance. The frequency-domain response matrix
$\boldsymbol{H}_n$ can be obtained as
\begin{align}
\boldsymbol{H}_n=\sum^{D}_{d=1}\widehat{\boldsymbol{H}}_{d}\exp\left(\frac{-j2\pi
dn}{N}\right).
\end{align}
For the FITRA algorithm, the regularization parameter is set to be
$\lambda=0.25$ as suggested by \cite{StuderLarsson13}. Also,
unless explicitly stated otherwise, the maximum number of
iterations of the FITRA and the EM-TGM-GAMP are set to be 2000 and
200, respectively.

\begin{figure*}[!t]
\setlength{\abovecaptionskip}{0pt}
\setlength{\belowcaptionskip}{0pt} \centering
\includegraphics[width=17cm]{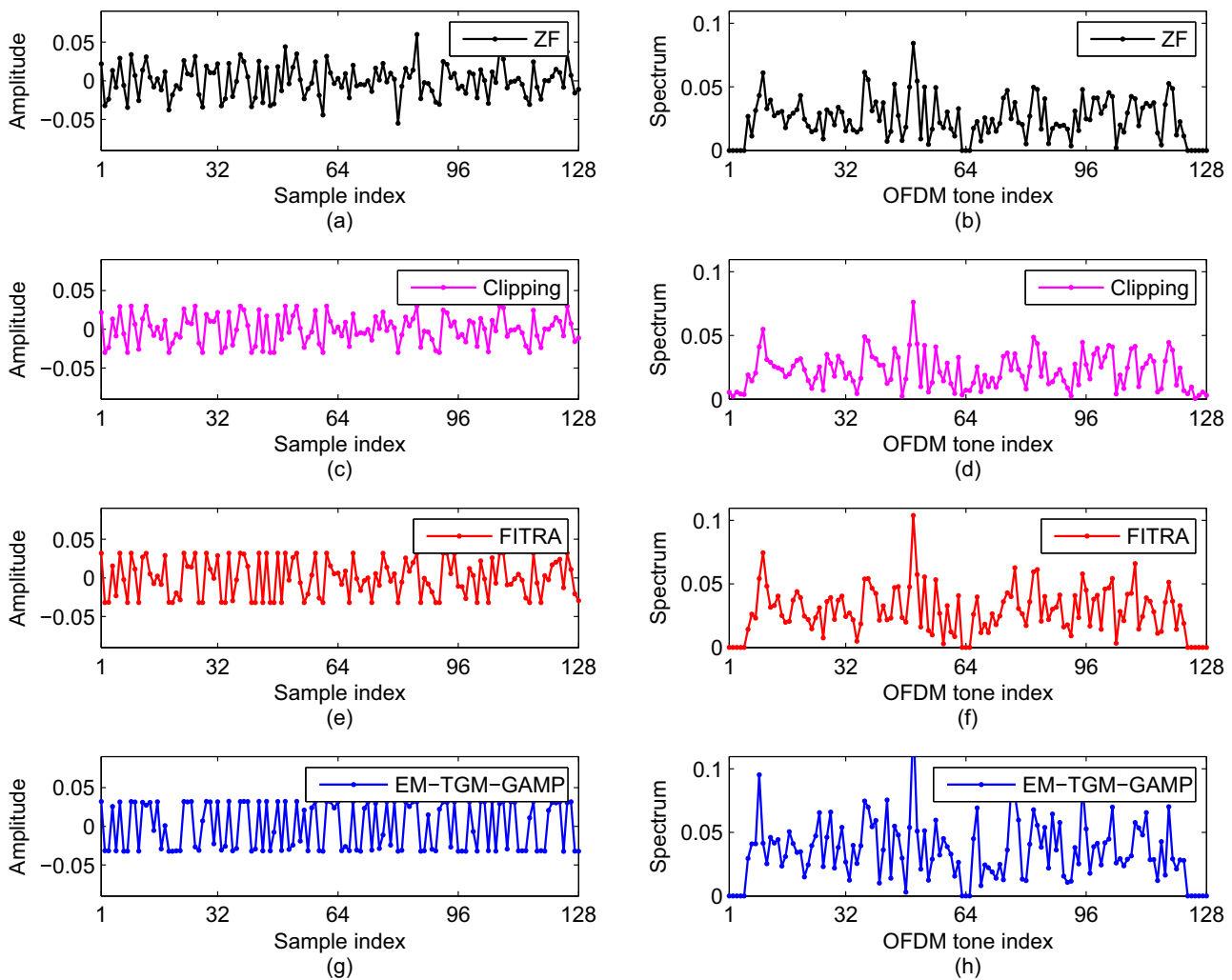}
\caption{ Time/Frequency representation for different schemes.
(a), (c), (e) and (g) are time-domain signals for ZF, clipping,
FITRA and EM-TGM-GAMP, respectively (\textsf{PAPR}:
$\text{ZF}=10.6\,$dB, $\text{Clipping}=4.3\,$dB,
$\text{FITRA}=2.4\,$dB, and $\text{EM-TGM-GAMP}=0.8\,$dB). (b),
(d), (f) and (h) are frequency-domain signals for respective
schemes (\textsf{MUI}: $\text{ZF}=-\infty\,$dB,
$\text{Clipping}=-15.3\,$dB, $\text{FITRA}=-64.1\,$dB, and
$\text{EM-TGM-GAMP}=73.6\,$dB; \textsf{OBR}:
$\text{ZF}=-\infty\,$dB, $\text{Clipping}=-13.8\,$dB,
$\text{FITRA}=-60\,$dB, and $\text{EM-TGM-GAMP}=-70.5\,$dB) }
\label{signal}
\end{figure*}

The complementary cumulative distribution
function (CCDF) is used to evaluate the PAPR reduction
performance. The CCDF denotes the probability that the PAPR of the
estimated signal exceeds a given threshold $\textsf{PAPR}_0$, i.e.
\begin{align}
\textsf{CCDF}(\textsf{PAPR}_0)=Pr(\textsf{PAPR}>\textsf{PAPR}_0).
\end{align}
Also, to evaluate the multiuser interference of the transmit signals, we define the $\textsf{MUI}$ as
\begin{align}
\textsf{MUI}=\frac{\sum_{n\in\mathcal{T}}\|\boldsymbol{s}_n-\boldsymbol{H}_n\boldsymbol{w}_n\|^2_2}
{\sum_{n\in\mathcal{T}}\|\boldsymbol{s}_n\|^2_2}.
\end{align}
Besides, the out-of-band (power) ratio (OBR) is introduced to measure
the out-of-band radiation of the solution, which is define as
\begin{align}
\textsf{OBR}=\frac{|\mathcal{T}|\sum_{n\in\mathcal{T}^{c}}\|\boldsymbol{w}_n\|^2_2}
{|\mathcal{T}^{c}|\sum_{n\in\mathcal{T}}\|\boldsymbol{w}_n\|^2_2}.
\end{align}
Note that, for the ZF procoding scheme, we have $\textsf{OBR}=0$
and $\textsf{MUI}=0$, while for the other three schemes, we always
have $\textsf{OBR}>0$ and $\textsf{MUI}>0$.

It is interesting to examine the signals estimated by respective
schemes. In the (a), (c), (e) and (g) of Fig. \ref{signal}, we
depict the real-part of the first transmit antenna's time-domain
signal (i.e. $\boldsymbol{\hat{a}}_1$) estimated by respective
schemes (the imaginary part behaves similarly). We observe that
our proposed algorithm yields a solution with most of its entries
(about $84.4\%$) located on the boundary points, which
corroborates our previous claim that the proposed truncated
hierarchical Gaussian mixture model encourages a quasi-constant
magnitude solution. Such a solution, clearly, has a low PAPR as it
looks like a constant-modulus signal. The solution of the FITRA
algorithm has fewer entries (about $49.2\%$) located on the
boundary points. For the ZF scheme, its solution exhibits a large
variation with a few high peaks. The solution of the clipping
scheme is only a slightly alleviated version of the ZF solution.
Numerical results also verify our observations: our proposed
algorithm has the lowest PAPR (PAPR associated with the first
transmit antenna) of $0.8\,$dB, the FITRA algorithm and the
clipping scheme have higher PAPRs of $2.4\,$dB and $4.3\,$dB,
repectively, while the ZF scheme has the highest PAPR of
$10.6\,$dB. We see that our proposed algorithm renders a much
lower PAPR than the other three schemes. The (b), (d), (f) and (h)
of Fig. \ref{signal} depict the magnitudes of the frequency-domain
signal $\boldsymbol{a}_1$ vs. the OFDM tone index. Both the
EM-TGM-GAMP and the FITRA have small MUIs and out-of-band
radiations: their MUIs are given by $-73.6\,$dB and $-64.1\,$dB,
respectively, and OBRs are given by $-70.5\,$dB and $-60\,$dB,
respectively. In contrast, the clipping scheme incurs a much
higher MUI and out-of-band distortion, with its MUI and OBR given
by $-15.3\,$dB and $-13.8\,$dB, respectively.


\begin{figure*}[!t]
\setlength{\abovecaptionskip}{0pt}
\setlength{\belowcaptionskip}{0pt} \centering
\includegraphics[width=17cm]{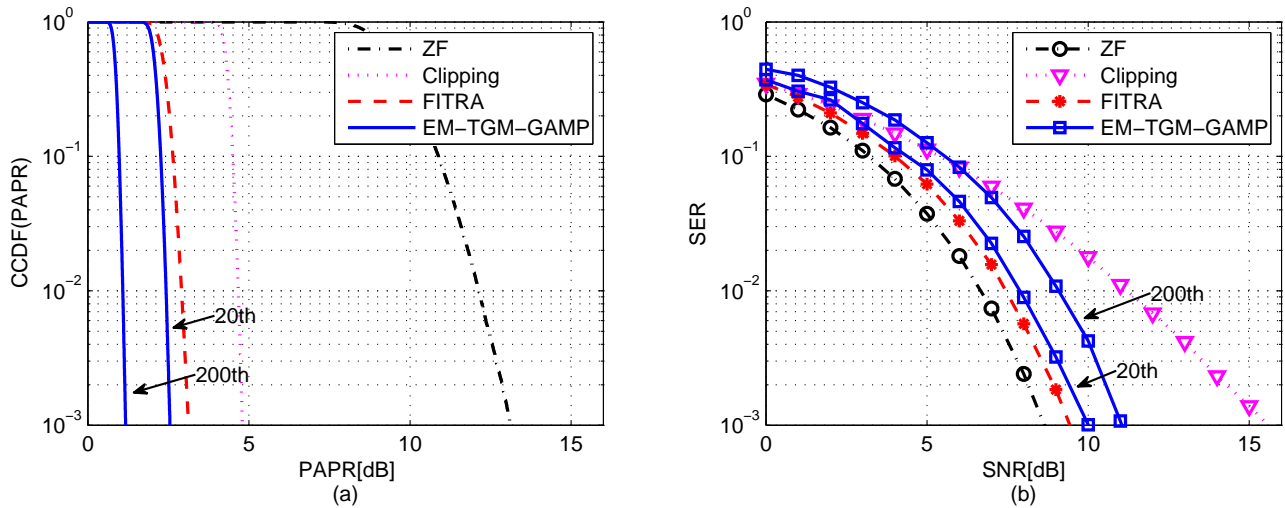}
\caption{PAPR and  symbol error rate (SER) performance for various
schemes. (a) CCDF of the PAPR, (b) SER performance.}
\label{PAPR_BER}
\end{figure*}

To better evaluate the PAPR reduction performance, we plot the
CCDF of the PAPR for respective schemes in Fig. \ref{PAPR_BER}(a).
The number of trials is chosen to be 1000 in our experiments. Note
that PAPRs associated with all $M$ antennas are taken in account
in calculating the empirical CCDF. We also include the results of
our proposed algorithm obtained at the $20$th iteration. We can
see that our proposed algorithm with 200 iterations achieves a
substantial PAPR reduction: it reduces the PAPR by more than
$11\,$dB compared to the ZF scheme (at $\textsf{CCDF(PAPR)}=1\%$),
by about $2\,$dB compared to the FITRA algorithm with $2000$
iterations, and by about $3.2\,$dB compared to the clipping
scheme. Also note that the proposed algorithm with only $20$
iterations can obtain a PAPR that even is lower than the FITRA,
meanwhile exhibiting a decent MUI and OBR (here MUI and OBR are
averaged over 1000 independent runs) given by $-41.8\,$ dB and
$-21.7\,$ dB, respectively.

The SER performance of respective schemes is shown in Fig.
\ref{PAPR_BER}(b), where the signal-to-noise ratio (SNR) is
defined as $\textsf{SNR}=\|\boldsymbol{x}\|_2^2/MN_o$, $N_o$
denotes the variance of the receiver noise (c.f.
(\ref{transmition})). We observe that the proposed algorithm
incurs an SNR-performance loss of $2.5\,$dB and $1.7\,$dB (at
$\textsf{SER}=10^{-3}$) compared to the ZF and FITRA schemes,
respectively. This performance loss, as discussed in
\cite{StuderLarsson13}, is primarily due to an increase in the
norm of the obtained solution $\boldsymbol{x}$, i.e.
$\|\boldsymbol{x}\|_2^2$. It is not difficult to see that the
solution obtained by our proposed method has a larger norm than
the solution of the FITRA since our solution has more entries
located on the boundary points. Also note that the ZF scheme
renders the least-norm solution. In order to maintain the same
SNR, our solution requires a stronger normalization, which causes
the SER performance loss compared to the ZF and FITRA schemes. It
can also be observed that the SER performance gap can be reduced
if we only perform $20$ iterations for our proposed method, in
which case the resulting solution has fewer entries located on the
boundary points and hence the increase of the norm of the solution
is not that significant. Note that the performance loss of the
clipping scheme is mainly caused by the residual MUI.

\begin{figure}[!t]
\setlength{\abovecaptionskip}{0pt}
\setlength{\belowcaptionskip}{0pt} \centering
\includegraphics[width=7.9cm]{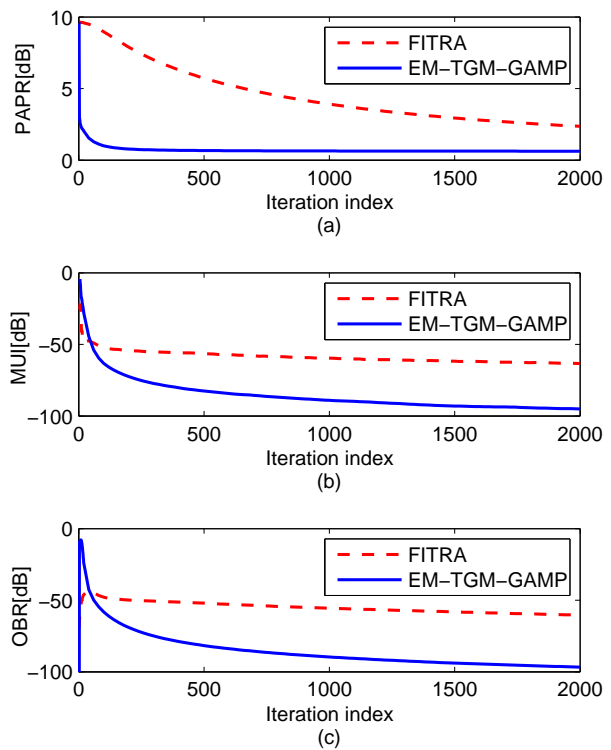}
\caption{Convergence rates of different metrics for EM-TGM-GAMP
and FITRA. (a) PAPR, (b) MUI, (c) OBR } \label{speed}
\end{figure}

We now examine the convergence rates of our proposed method and
the FITRA algorithm. The (a), (b) and (c) of Fig. \ref{speed} show
the PAPR, MUI and OBR vs. the number of iterations, respectively.
Results are averaged over 1000 independent runs and the PAPR
results are averaged over PAPRs associated with all transmit
antennas. Our numerical results show that the average MUI and OBR
of our proposed method at the $200$th iteration are $-72.5\,$dB
and $-69.1\,$dB respectively, while the average MUI and OBR of the
FITRA algorithm at the $2000$th iteration are $-63.3\,$dB and
$-60.3\,$dB. With less than $200$ iterations, our proposed
algorithm achieves better MUI cancelation than the FITRA algorithm
with even $2000$th iterations. From Fig. \ref{speed}(a), we also
notice that our proposed algorithm has a fast convergence rate and
is able to obtain a low-PAPR solution within only 200 iterations,
whereas it takes the FITRA algorithm about 2000 iterations to
achieve a reasonably low PAPR.



\begin{figure}[!t]
\setlength{\abovecaptionskip}{0pt}
\setlength{\belowcaptionskip}{0pt} \centering
\includegraphics[width=7.9cm]{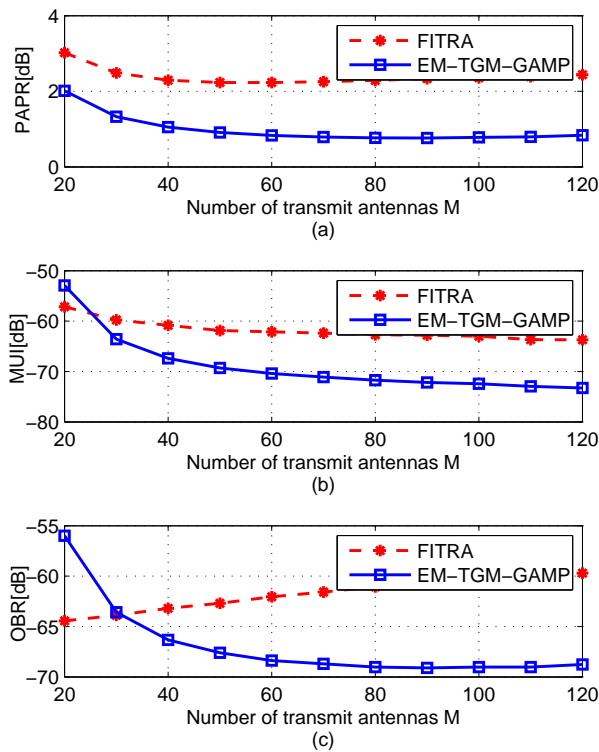}
\caption{(a) PAPR vs. number of transmit antennas, (b) MUI vs. number of transmit antennas,
(c) OBR vs. number of transmit antennas.} \label{varyAntenna}
\end{figure}

Lastly, we investigate the PAPR-reduction performance under
different number of transmit antennas, where the number of users
is fixed to be $K=10$, and the number of transmit antennas at the
BS varies from $20$ to $120$. Fig. \ref{varyAntenna} plots the
PAPR, MUI and OBR as the number of transmit antennas varies, where
results are averaged over 1000 independent runs and the PAPR
results are averaged over PAPRs associated with all transmit
antennas. We observe that both algorithms achieve a low PAPR when
sufficient DoFs at the base station are available. Nevertheless,
the proposed method is capable of exploiting the available DoFs
more efficiently as the number of of transmit antennas increases.

\section{Conclusions} \label{sec:conclusion}
We considered the problem of joint PAPR reduction and multiuser
interference (MUI) cancelation in OFDM based massive MIMO downlink
systems. A hierarchical truncated Gaussian mixture prior model was
proposed to encourage a low PAPR solution/signal. A variational EM
algorithm was developed to obtain estimates of the hyperparameters
associated with the prior model, as well as the signal.
Specifically, the GAMP technique was embedded into the variational
EM framework to facilitate the algorithm development. The proposed
algorithm only involves simple matrix-vector multiplications at
each iteration, and thus has a low computational complexity.
Simulation results show that the proposed algorithm achieves
notable improvement in PAPR reduction as compared with the FITRA
algorithm \cite{StuderLarsson13}, and meanwhile renders better MUI
cancelation and lower out-of-band radiation. The proposed
algorithm also demonstrates a fast convergence rate, which makes
it attractive for practical real-time systems.

\bibliography{newbib}
\bibliographystyle{IEEEtran}

\end{document}